\input harvmac
\input psfig
\newcount\figno
\figno=0
\def\fig#1#2#3{
\par\begingroup\parindent=0pt\leftskip=1cm\rightskip=1cm\parindent=0pt
\global\advance\figno by 1
\midinsert
\epsfxsize=#3
\centerline{\epsfbox{#2}}
\vskip 12pt
{\bf Fig. \the\figno:} #1\par
\endinsert\endgroup\par
}
\def\figlabel#1{\xdef#1{\the\figno}}

\def\encadremath#1{\vbox{\hrule\hbox{\vrule\kern8pt\vbox{\kern8pt
\hbox{$\displaystyle #1$}\kern8pt}
\kern8pt\vrule}\hrule}}
\def\underarrow#1{\vbox{\ialign{##\crcr$\hfil\displaystyle
 {#1}\hfil$\crcr\noalign{\kern1pt\nointerlineskip}$\longrightarrow$\crcr}}}
%
\overfullrule=0pt

%

\def\R{{\bf R}}

\font\zfont = cmss10 

\def\bigone{\hbox{1\kern -.23em {\rm l}}}
\def\ZZ{\hbox{\zfont Z\kern-.4emZ}}

\Title{hep-th/0112258} {\vbox{\centerline{Multi-Trace Operators,
Boundary Conditions,}
\bigskip
\centerline{And AdS/CFT Correspondence }}}
\smallskip
\centerline{Edward Witten}
\smallskip
\centerline{\it Institute For Advanced Study, Princeton NJ 08540 USA}


\def\O{{\cal O}}
\medskip
\vskip 1cm
 \noindent
 We argue that multi-trace interactions in  quantum
 field theory on the boundary of AdS space
 can be incorporated in the AdS/CFT correspondence by using
 a more general boundary condition for the bulk fields than has
 been considered hitherto.  We illustrate the procedure for
 a renormalizable four-dimensional field theory with a
 $(\Tr\,\Phi^2)^2$ interaction.  In this example, we show how the AdS
 fields with the appropriate boundary condition reproduce the
 renormalization group effects found in the boundary field theory.
 We also construct in  related examples a line of fixed points
 with a nonperturbative duality and a flow between two methods of
 quantization.

 \Date{December, 2001}

\newsec{Introduction}

Large $N$ limits for matrix-valued fields have been a subject of
considerable interest ever since the large $N$ limit of $SU(N)$
gauge theory was recognized as a likely starting point for
understanding the dynamics of four-dimensional quantum gauge
theories \ref\thooft{G. 't Hooft, ``A Planar Diagram Theory For
Strong Interactions,'' Nucl. Phys. {\bf B72} (1974) 461.}. Much
recent work centers around the AdS/CFT correspondence
\ref\malda{J. Maldacena, ``The Large $N$ Limit Of Superconformal
Field Theories and Supergravity,'' Adv. Theor. Math. Phys. {\bf
2} (1998) 231. }.

In general, given a collection of matrix-valued fields $\Phi_i$,
to construct a theory with a large $N$ limit, one considers
normalized trace operators
 \eqn\hudo{\O_\alpha={1\over N}
\Tr\,F_\alpha(\Phi_i)}
 and an action functional
\eqn\nudo{I=N^2W(\O_\alpha).}
 Here, the $F_\alpha$ are arbitrary
functions\foot{In many applications, one considers only polynomial
functions.} of the $\Phi_i$ and their derivatives, and $W$ is an
arbitrary function of the $\O_\alpha$.  $F_\alpha$ has no explicit
dependence on $N$ and is defined without any traces, so
$\O_\alpha$ is a ``single-trace operator,'' and $W$ likewise has
no explicit dependence on $N$. The powers of $N$ have been chosen
to ensure the existence of a large $N$ limit.

\nref\prek{S. R. Das, A.
Dhar, A. M. Sengupta, and S. R. Wadia, ``New Critical Behavior in $D=0$
Large $N$ Matrix Models,'' Mod. Phys. Lett. {\bf A5} (1990) 1041.}%
\nref\crna{L. Alvarez-Gaum\'e, J. L. Barbon, and C. Crnkovic, ``A
Proposal For Strings At $D>1$,'' Nucl. Phys. {\bf B394} (1993)
383.}%
\nref\korch{G. Korchemsky, ``Matrix Model Perturbed By Higher Order Curvature
Terms,'' Mod. Phys. Lett. {\bf A7} (1992) 3081,  ``Loops In The Curvature
Matrix Model,'' Phys. Lett. {\bf B296} (1992) 323.}%
 \nref\kleb{I. R. Klebanov, ``Touching
Random Surfaces And Liouville Gravity,'' Phys. Rev. {\bf D51}
(1995) 1836; I. R. Klebanov and A. Hashimoto, ``Nonperturbative
Solution Of Matrix Models Modified By Trace-Squared Terms,''
Nucl. Phys. {\bf B434} (1995) 264;  J. L. F. Barbon, K.
Demeterfi, I. R. Klebanov, and C. Schmidhuber, ``Correlation
Functions In Matrix Models Modified By
Wormhole Terms,'' Nucl. Phys. {\bf B440} (1995) 189.}%
\nref\kh{I. R. Klebanov and A. Hashimoto, ``Wormholes, Matrix
Models, and Liouville Gravity,'' in {\it String Theory, Gauge
Theory, and Quantum Gravity}, ed. R. Dijkgraaf et. al. (North
Holland, 1996).}%
 \nref\abs{O. Aharony, M. Berkooz, and E. Silverstein,
``Multiple Trace Operators And Nonlocal String Theories,'' JHEP
{\bf 0108:006} (2001), hep-th/0105309, ``Non-Local String Theories
On $AdS_3\times S^3$ And Stable Non-Supersymmetric Backgrounds,''
hep-th/0112178.}%
 If $W$ is a
linear function of the $\O_\alpha$'s, then it is a ``single-trace
action,'' while nonlinear terms in $W$ are ``multi-trace
interactions.''  Many important examples, like four-dimensional
gauge theory without matter fields, are based on single-trace
actions.  Theories with multi-trace interactions have, however,
also been considered in matrix models of two-dimensional gravity
\refs{\prek -\kh} and in the AdS/CFT correspondence \refs{\abs}.

\nref\gkp{S. Gubser, I. Klebanov, and A. M. Polyakov, ``Gauge Theory
Correlators From Noncritical String Theory,'' Phys. Lett. {\bf B428}
(1998) 105, hep-th/9802109.}%
 \nref\witten{E. Witten, ``Anti De Sitter Space And Holography,''  Adv.
 Theor. Math. Phys. {\bf 2} (1998) 253, hep-th/9802150.   }%
The purpose of the present paper is to make a general proposal for
how multi-trace interactions can be incorporated in the AdS/CFT
correspondence, by generalizing the boundary conditions
\refs{\gkp,\witten} that are used in the case of single-trace
actions.  This is somewhat subtle for reasons explained in
\refs{\abs}.  Multi-trace operators in the CFT correspond to
multi-particle states in AdS space, and it is not immediately
apparent what is meant by placing a boundary condition at
infinity on a multi-particle state.  Nevertheless, we will see
that the familiar boundary conditions have a natural
generalization that serves this purpose.

In section 2, we practice by recalling  how multi-trace
interactions are treated in the simplest of all  matrix models,
which is the theory of a single matrix. This model can be solved
very directly in  the large $N$ limit \ref\brezin{E. Brezin, C.
Itzykson, G. Parisi, and J.-B. Zuber, ``Planar Diagrams,''
Commun. Math. Phys. {\bf 59} (1978) 35.} and the solution can be
extended to incorporate multi-trace interactions \prek.\foot{This
theory also has scaling limits related to two-dimensional gravity
that can be solved in a much more subtle way \ref\gdb{D. J. Gross
and A. A. Migdal, Phys. Rev. Lett. {\bf 64} (1990) 717; M. Douglas
and S. Shenker, Nucl. Phys. {\bf B335} (1990) 635; E. Brezin and
V. Kazakov, Phys. Lett. {\bf 236B} (1990) 144.}, but we do not
need that here.} In section 3, we present our proposal for the
general boundary conditions in the AdS/CFT correspondence and
their interpretation in terms of multi-trace interactions. In
section 4, we discuss the application of this proposal to a class
of four-dimensional examples suggested in \abs.  We also describe
in related examples a line of fixed points that admits a
nonperturbative duality.

The role of boundary conditions in the AdS/CFT correspondence has
been re-examined recently in \ref\mr{P. Minces and V. O. Rivelles,
``Energy And The AdS/CFT Correspondence,'' hep-th/0110189.}. The
paper \ref\berkooz{M. Berkooz, A. Sever, and A. Shomer, ``
`Double-Trace' Deformations, Boundary Conditions, And Spacetime
Singularities,'' hep-th/0112264.}, which appeared on hep-th the
same day as the present one, has some results that overlap with
those presented here.

\newsec{Review Of The One-Matrix Model}

First we practice by recalling the case of a model (in zero
space-time dimensions) in which the field variables comprise a
single $N\times N$ hermitian matrix $\Phi$.  Assuming that we
want to consider only polynomial interactions (in this particular
model, there is no difficulty in relaxing this assumption), the
general single-trace operators are
 \eqn\normo{\O_n={1\over N}\Tr\,\Phi^n,~~n=1,2,3,\dots.}
 The action depends on an arbitrary function of the $\O_n$'s:
 \eqn\kormo{I=N^2W(\O_1,\O_2,\dots).}
One wishes to perform the integral
 \eqn\tormo{Z=\int d\Phi \,\exp(-I)}
 in the large $N$ limit, as a function of $W$.

   The model has $U(N)$ symmetry, acting on $\Phi$
 by conjugation.  Because of this symmetry, the action depends only on the
 eigenvalues $\lambda_1,\dots,\lambda_N$ of $\Phi$; in fact,
 $\O_n={1\over N}\sum_i\lambda_i^n$.
Up to a constant factor (computable and independent of $W$) which
comes from the volume of the group $U(N)$, the partition function
in terms of the eigenvalues becomes
  \eqn\thormo{Z=\int_{-\infty}^\infty d\lambda_1\dots d\lambda_N
  \prod_{i<j}(\lambda_i-\lambda_j)^2\exp(-N^2W).}
  The integral can be evaluated for large $N$ because the
  integrand has a sharp maximum at a suitable configuration of the
  $\lambda$'s. This configuration is characterized by a smooth distribution of
  the $\lambda$'s.   To find it, we assume that the
  density of the $\lambda$'s is of the form $N\rho(x)$ for some
  positive function $\rho(x)$ with $\int_{-\infty}^\infty dx
  \,\rho(x)=1$.
The $\O_n$ now become $\O_n=\int_{-\infty}^\infty  dx\,\,
x^n\rho(x)$, and the integrand in \thormo\ is
 \eqn\jormo{\exp\left(-N^2( W(\O_i)-\int dx\,dx'\rho(x)\rho(x')\ln
 |x-x'|)\right).}
Maximizing the exponent with the constraint $\int_{-\infty}^\infty
dx \,\rho(x)=1$ gives the equation
 \eqn\utormo{\sum_nx^n{\partial W\over \partial \O_n} -2\int
 dx'\ln|x-x'| \rho(x') +t =0,}
  where $t$ is a Lagrange multiplier for the constraint.
 How to solve such equations, for suitable $W$'s, is explained in \brezin.

Let us consider what happens in the single-trace case, for which
$W=\sum_{n=1}^\infty w_n\O_n$ with constants $w_n$. Then \jormo\
simply reduces to
 \eqn\vutormo{\sum_nx^nw_n -2\int
 dx'\ln|x-x'| \rho(x') +t =0.}
The lesson to be learned is that in the general case, there is a
saddle-point equation of the same form as in the single-trace
case, except that the coupling constants $w_n$ of the
single-trace case are replaced by $\partial W/\partial\O_n$,
where $\O_n$ are the observables of the matrix field theory under
study and $W$ is the coupling function which can be, in general,
an arbitrary function of these observables.

\newsec{The AdS/CFT Case}

The remainder of this paper is concerned mainly with adapting that
last statement to the AdS/CFT correspondence.  However, first we
practice with another familiar example of a different type.

Consider a scalar field $\phi$ on the half-space $x_1\geq 0$ in
$\R^n$.  We take the action for $\phi$ to be a sum
 \eqn\hugot{I={1\over 2}\int_{x_1\geq 0} d^nx \,|d\phi|^2
 +\int_{x_1=0}d^{n-1}x \,\,W(\phi,d\phi,\dots)}
 of the free kinetic energy for $\phi$, integrated over the
 half-space, plus a boundary interaction $W$
that is an arbitrary local function of $\phi$ and its derivatives.
Now when we vary $\phi$, $\phi\to \phi+\delta \phi$, to obtain
the Euler-Lagrange equations, we encounter boundary terms
 \eqn\tutter{\int_{x_1=0}d^{n-1}x\,\delta\phi\left(-{\partial
 \phi\over \partial x_1}+{\delta W\over \delta \phi}\right).}
 Hence, to satisfy the equations of motion, we have to impose a
 boundary condition
\eqn\mutter{{\delta W\over \delta \phi}-{\partial
\phi\over\partial x_1}=0.}

This equation has an analogy with \utormo.  The term $\delta
W/\delta \phi$ is analogous to the term $\sum_nx^n{\partial W\over
\partial \O_n}$ in \utormo, while $-\partial\phi/\partial x_1$
corresponds to the terms in \utormo\ that do not depend on $W$.

\mutter\ can also be given the following intuitive interpretation.
Let us think of $x_1$ as a ``time'' direction, even though in a
boundary problem like this one with local boundary conditions
(and also in the AdS case that we turn to presently) it is
usually a spatial direction.  Then $p=\partial\phi/\partial x_1$
is naturally regarded as the ``momentum.''  For $W=0$, the
boundary condition is simply the vanishing of the momentum,
$p=0$.  For  general $W$, the boundary condition $p=\delta
W/\delta \phi$  differs by a canonical transformation (generated
by $W$) from the condition $p=0$.  In the phase space of
canonically conjugate variables $\phi $ and $p$ (defined at
$x_1=0$), the variables $p$ or equally well $p-\delta W/\delta
\phi$ are a maximal set of commuting variables, so their
vanishing defines a ``Lagrangian submanifold'' of the phase
space. A Lagrangian submanifold determines a quantum state at
least formally (in the present case, the operators $p-\delta
W/\delta \phi$ annihiliate the state with wave function $e^{-W}$).
In two-dimensional conformal field theory, the state determined in
this way by the boundary conditions is called the boundary state
\ref\cn{C. G. Callan, Jr., C. Lovelace, C. R. Nappi, and S. A.
Yost, ``Adding Holes And Crosscaps To The Superstring,'' Nucl.
Phys. {\bf B293} (1987) 83.}.

\bigskip\noindent{\it Analog for Anti de Sitter Space}

Anti de Sitter space is somewhat similar, except that the
boundary is replaced by a conformal boundary at spatial infinity.
We consider AdS space of dimension $D=d+1$, with metric
 \eqn\adsmet{ds^2={dr^2+\sum_{i=1}^ddx_i^2\over r^2}}
 in the region $r\geq 0$.
 The conformal boundary is at $r=0$.

 Consider a scalar field $\phi$ of mass $m$ in AdS space.  It
 behaves near $r=0$ as \eqn\gsca{\phi=\alpha(x) r^{d-\lambda}
 +\beta(x) r^\lambda,} where  we take $\lambda$ to be the larger
 root of the equation $\lambda(\lambda+d)=m^2$.  (For
 $\lambda=d/2$, a case we consider in section 4, the roots are
 equal and the two solutions are $r^{d/2}$ and $r^{d/2}\ln r$.)
 $\alpha$ and $\beta$ are canonically conjugate variables,
 analogous to $\phi$ and $\partial_1\phi$ in the example treated
 above in which  the boundary is at finite distance.

\nref\freedetal{D. Z. Freedman and
 P. Breitenlohner and D. Z. Freedman, ``Stability In Gauged Extended
 Supergravity,'' Ann. Phys. {\bf 144} (1982) 249.}%
 \nref\ha{V. Balasubramanian, P. Kraus, and A. Lawrence, ``Bulk vs.
Boundary Dynamics In Anti-de Sitter Spacetimes,'' hep-th/9808017.}%
 \nref\kw{I. Klebanov and E. Witten, ``AdS/CFT Correspondence And Symmetry
 Breaking,'' Nucl. Phys. {\bf B556} (1999) 89, hep-th/9905104.}%

 In the AdS/CFT correspondence, one interprets $\beta(x)$ as the
 expectation value of a scalar field $\O$ of dimension $\lambda$
 in the boundary conformal field theory.\foot{For a certain range
 of negative values of $m^2$, there are two ways to quantize the
 $\phi$ field in AdS space \freedetal, and correspondingly $\O$
 can have dimension $d-\lambda$
instead of $\lambda$ \refs{\ha,\kw}.  This possibility will enter
in section 4.}
 $\alpha(x)$ is related in a way we specify presently to a source
 for $\O$.  The relationship between the fields and sources is
 analogous to the relationship between normalizable and unnormalizable
 operators in Liouville theory \ref\polly{N. Seiberg, ``Notes On Quantum
 Liouville Theory And Quantum Gravity,'' Prog. Theor. Phys. Suppl.
 {\bf 102} (1990) 319.}.

 For theories with the familiar sort of large $N$ limit, $\O$ is a
 single-trace operator.  To compute the expectation value of
 $\exp(-N^2\int d^nx f(x) \O)$ in the boundary conformal field theory,
 the familiar recipe \refs{\gkp,\witten} is to compute the AdS
 partition function with the boundary condition
 \eqn\jungo{\alpha=f}
 on $\phi$.

Computing the expectation value of $\exp(-N^2\int d^nx \,f(x)\O)$
is the same as computing the partition function of the boundary
conformal field theory in the presence of a perturbation $N^2W$
added to the Lagrangian, where $W=\int d^nx \,f(x)\O$.  Since
$\beta$ corresponds in the AdS/CFT correspondence to the
expectation value of $\O$, we can symbolically write the boundary
coupling as $W=\int d^nx\, f\beta$.\foot{The factor of $N^2$
multiplying $W$ is replaced in the string theory by a factor of
$1/g_s^2$, where $g_s$ is the string coupling constant.  Both the
bulk Lagrangian and the boundary coupling have this factor, so it
does not show up in the boundary condition.}
 If we do
this, then the boundary condition \jungo\ can be written
 \eqn\tungo{\alpha={\delta W\over \delta \beta}.}

The problem of multi-trace interactions arises if we replace $W$
by a local but nonlinear functional $W(x,\O,d\O,\dots)$ of $\O$
and its derivatives.  Now we can state our proposal for
incorporating multi-trace interactions in the AdS/CFT
correspondence: interpret $W$ as a functional of $\beta$ by
replacing $\O$ everywhere with $\beta$ to get a functional
$W(x,\beta,d\beta,\dots)$ and impose the boundary condition
\tungo\ whether $W$ is linear or not.

This generalizes immediately to the case of several scalar fields
$\phi_i$ of masses $m_i$.  They behave near $r=0$ as
 \eqn\gsca{\phi_i=\alpha_i(x)
r^{d-\lambda_i} +\beta_i(x) r^{\lambda_i}.} The $\beta_i$ are
related to expectation values of operators $\O_i$ of dimension
$\lambda_i$ in the boundary field theory, and the $\alpha_i$ are
related to sources for those operators.  Given a general
multi-trace interaction $W(x,\O_i,d\O_i,\dots)$ in the boundary
theory, to incorporate it in the bulk theory we impose the
boundary condition \eqn\sca{\alpha_i={\delta
W(x,\beta_k,d\beta_k,\dots)\over \delta \beta_i}.}

\newsec{$(\Tr\,\Phi^2)^2$ Interaction In Four Dimensions And Related Examples}

To illustrate this proposal, we consider first an example along
lines suggested in \abs: a renormalizable field theory in four
dimensions with a double-trace interaction.

Let  $\O$ be a half-BPS operator of dimension 2 in, for example,
${\cal N}=4$ super Yang-Mills theory in four dimensions.  (For
example, take $\O=\Tr\,(\Phi_1^2-\Phi_2^2)$ where $\Phi_1$ and
$\Phi_2$ are two of the scalar fields.)  We want to perturb the
theory with the boundary coupling
 \eqn\huy{W={f\over 2}\int d^4x \,\O^2.}
 $f$ is a dimensionless coupling constant. Classically,
turning on $f\not= 0$ preserves conformal invariance while
completely breaking supersymmetry.  Quantum mechanically,
conformal invariance is violated in order $f^2$.

In fact, on flat $\R^4$ at $f=0$, the two-point function of the
operator $\O$   is determined by conformal invariance to be
$\langle \O(x)\O(y)\rangle = v/|x-y|^4$, with $v>0$ by
unitarity.  This leads to a non-trivial beta function in order
$f^2$. Indeed, to compute quantum mechanical amplitudes in order
$f^2$, we would need matrix elements of
 \eqn\needma{{f^2\over 8}\int d^4x\,
 d^4y\, \O^2(x)\O^2(y).}
To evaluate the divergent contributions in \needma, we need to
know the operator product expansion of $\O(x)\O(y) $ for $x\to y$.
In the large $N$ limit, the structure simplifies drastically: the
two factors of $\O$ in $\O^2(x)$ or $\O^2(y)$ do not
``interfere'' with each other.   A divergent term that
renormalizes $f$ and survives in the large $N$ limit comes only
from the identity operator appearing in the product $\O(x)\O(y)$
for one pair of $\O$'s (analogous to a ``single contraction'' in
free field theory).  So we have to evaluate the divergent part of
 \eqn\needb{{f^2\over 2}\int d^4xd^4y\,\O(x)\O(y) \langle
 \O(x)\O(y)\rangle.}    Setting $w=y-x$, we
 encounter the logarithmically divergent integral \eqn\heedb{\int
 d^4w \O(x)\O(x+w)\langle\O(0)\O(w)\rangle
  \sim \O(x)^2\int d^4w\,{v\over |w|^4}\sim 2\pi^2v\ln\Lambda\cdot
  \O(x)^2,} where $\Lambda$ is a cutoff.
This divergence, since it multiplies the operator $\O^2(x)$ that
appears in the original interaction \huy, can be interpreted as a
renormalization of $f$.  Because correlation functions factorize
in the large $N$ limit, there are no higher order corrections to
the beta function; the renormalization just described gives the
full answer. Henceforth, we normalize $\O$ so that the beta
function coefficient is 1.

Now, let us try to reproduce this behavior in the AdS language.
Because $\O$ has dimension $2=d/2$, the corresponding scalar
field $\phi$ in AdS space has an exceptional behavior
 \eqn\tuyf{\phi= \alpha(x) r^2\ln (\mu r)+\beta(x) r^2}
 near the boundary.  Here $\mu$ is an arbitrary scale factor that
 must be included to define the logarithm.  Our boundary
 interaction, interpreted in AdS language, is $W={f\over
 2}\int d^4x \,\beta^2$.  Hence, the boundary condition
 \tungo\ becomes
 \eqn\hungo{\alpha=f\beta.}
 With this boundary condition, the field behaves
 near infinity as
 \eqn\ungo{\phi=\beta \,r^2(f\ln(\mu r)+1).}

 Let us try to extract from this the renormalization effects of
 the boundary field theory.  We want to re-express \ungo\ in terms
 of a bare coupling $f_0$ defined at a cutoff scale
 $\Lambda>>\mu$, and a bare field $\beta_0$.  We expect $\beta$ to
 be related to $\beta_0$ by multiplicative renormalization,
 $\beta=F(f_0,\Lambda/\mu)\beta_0$.  Since the observable quantity
 $\phi$ of the bulk theory must be independent of the
 quantity $\mu$ that was introduced in defining $\alpha $ and $\beta$
 (or alternatively, it must be independent of the renormalization procedure
 used on the boundary), we want
 \eqn\unz{\beta_0(f_0\ln(\Lambda r)+1)=\beta(f\ln(\mu r)+1).} It
 follows that $\beta_0f_0=\beta f $ and
  \eqn\tunz{f={f_0\over 1+f_0\ln(\Lambda/\mu)}.}
  \tunz\ is the typical relation between the renormalized coupling
  and the bare coupling in a theory with only a ``one-loop''
beta function.  Note that $f>0$ is required for
 positivity of the boundary coupling.  As in analogous examples
 of large $N$ bosonic theories discussed in \ref\gn{D.
 J. Gross and A. Neveu, ``Dynamical Symmetry Breaking In
 Asymptotically Free Field Theories,'' Phys. Rev. {\bf D10}
 (1974) 3235.}, the theory is not asymptotically free for
 $f>0$, though asymptotic freedom would arise formally for $f<0$.

\bigskip\noindent{\it A Line Of Fixed Points And A Nonperturbative Duality}

 Thus, we have shown in a non-trivial example how the more
 general boundary condition of the bulk theory reproduces the
 behavior of the boundary field theory.  We will now consider
 another example that is suggested in part by a recent
 investigation in
 de Sitter space \nref\andetal{M. Spradlin, A. Strominger, and A.
 Volovich, ``Les
 Houches
 Lectures On De Sitter Space,'' hep-th/0110007.}%
 \nref\poly{ R. Bousso, A. Maloney, and
 A. Strominger, ``Conformal Vacua And Entropy In De Sitter Space,''
 hep-th/0112218. }%
 \refs{\andetal,\poly}.  We
 consider in $D=d+1$-dimensional AdS space a theory with two scalar fields
 $\phi_1,\phi_2$ of equal mass squared.  Thus near the boundary
 \eqn\nearb{\phi_i\sim \alpha_i(x) r^{d-\lambda} +\beta_i(x)
 r^\lambda}
 with equal $\lambda$.  We suppose that $\lambda$ is in the range
 $d/2>\lambda>d/2-1$ where \freedetal\ two methods of quantization
 are possible.  We adopt one method of quantization for $\phi_1$
 and the second for $\phi_2$, so that \refs{\ha,\kw} $\phi_1$ is related in
 the boundary theory to a conformal primary operator $\O_1$ of dimension $\lambda$
 but $\phi_2$ is related to a conformal primary operator $\O_2'$ of dimension
 $d-\lambda$.   Because of the reversed method of quantizing
 $\phi_2$, we write $\beta_2=\alpha_2'$, $\alpha_2=\beta_2'$, so
 \nearb\ becomes
 \eqn\earb{\eqalign{\phi_1\sim \alpha_1(x) r^{d-\lambda} &+\beta_1(x)
 r^\lambda\cr \phi_i\sim \beta_2'(x) r^{d-\lambda} &+\alpha_2'(x)
 r^\lambda.\cr}}
Thus, $\beta_1$ and $\beta_2'$ are related to the expectation
values of $\O_1$ and $\O_2'$, and $\alpha_1$ and $\alpha_2'$ are
related to sources for those operators.

Now we want to perturb the boundary theory by the marginal
operator $f\O_1\O_2'$.
 This means that the
boundary functional is to be $W=f\int d^nx \beta_1\beta_2'$, and
hence the boundary condition \gsca\ is
 \eqn\uvu{\alpha_1=f\beta_2',~~\alpha_2'=f\beta_1.}
 The boundary condition preserves conformal invariance and is
 compatible with $\phi_1=\phi_2=0$.  Because $\phi_1$ and $\phi_2$
 are fields of non-zero mass squared, they vanish in the
 unperturbed AdS solution at $f=0$.  That solution still obeys the
 boundary conditions for $f\not=0$, and so if stable remains the correct
 solution for any $f$.  So (modulo stability)
  we get in this way a line of conformal
 fixed points, parameterized by $f$.
The dependence of this family on $f$ is non-trivial, since the
operator $\O_1\O_2'$ is a nonzero conformal primary, and
correlation functions computed using the boundary condition \uvu\
will certainly depend on $f$.

In fact, this line of fixed points admits a nonperturbative
duality. The boundary condition \uvu\  is invariant under $ f\to
1/f$ together with $\alpha_1\leftrightarrow \beta_1$,
$\alpha_2'\leftrightarrow \beta_2'$.  The latter operation is not
a symmetry of the full theory, since it is not a symmetry of the
$r\to 0$ formulas in \earb.  But suppose that the bulk theory has
a symmetry that exchanges $\phi_1$ and $\phi_2$.  Our method of
quantization broke this symmetry, because we quantized $\phi_1$
one way and $\phi_2$ the other way.  But
$\phi_1\leftrightarrow\phi_2$ is a symmetry when combined with
$f\leftrightarrow 1/f$. For this operation, which exchanges
$\alpha_1$ with  $\alpha_2=\beta_2'$ and $\beta_1$ with
$\beta_2=\alpha_2'$, is a symmetry of both \earb\ and \uvu\ as
well as being, by hypothesis, a symmetry of the bulk theory.

For $f\to\infty$, the physics is the same as at small $f$ except
that $\phi_1$ and $\phi_2$ are exchanged.  So in this limit,
$\phi_1$ is quantized to give an operator of dimension
$d-\lambda$ and $\phi_2$ to give an operator of dimension
$\lambda$, the reverse of the situation for small $f$.    Such a
switch in varying $f$ is possible because for $f$ of order one,
the two operators are mixed by the $\O_1\O_2'$ perturbation.

In the above, since we assumed $\lambda<d/2$, we could have added
a relevant perturbation $\O_1^2$, corresponding to a term
$\beta_1^2$ in $W$.  One can analyze the effects of such a
relevant perturbation in a fashion similar to the above.  In fact,
in doing so, let us for simplicity omit the field $\phi_2$ and
return to the case of a single scalar field $\phi$ with expansion
 \eqn\toggo{\phi\sim \alpha r^{d-\lambda}+
 \beta r^\lambda.}
 For $d/2>\lambda>d/2-1$, we can quantize the field with a
 boundary condition $\alpha=0$, in which case it corresponds to an
 operator $\O$ of dimension $\lambda$ in the boundary theory, or
 with a boundary condition $\beta=0$, in which case it corresponds
 to an operator $\O'$ of dimension $d-\lambda$ in the boundary
 theory.  Let us adopt the first method of quantization but
 include a relevant perturbation $W={g\over 2}\beta^2$.  The
 boundary condition is $\alpha=g\beta$, and we see that as
 $g\to\infty$, the boundary condition approaches the condition
 $\beta=0$ that is suitable for quantization to get a field of
 dimension $d-\lambda$.  So in fact, the renormalization group
 flow leads from one method of quantization to the other.  If we
 quantize with $\beta=0$ to get an operator of dimension
 $d-\lambda$, the perturbation that would probe this flow would be
  $W={g'\over 2}\alpha^2$, which is an irrelevant perturbation by
  an operator of dimension $2(d-\lambda)$.

The results we have just found are quite similar to results found
in the old matrix model \kleb, where adding a double trace
interaction with a suitable coefficient reverses the
``gravitational dressing'' of the interaction.  (For a review, see
\kh.)  Such a reversal is the Liouville analog of switching  from
$\lambda$ to $d-\lambda$ the dimension of the operator that is
associated with a given bulk field in the boundary theory.

\bigskip\noindent{\it A Remark On Boundary States}

In section 3, we noted that the boundary condition \gsca\ can be
interpreted formally as saying that the boundary condition
determines a quantum state, sometimes called the boundary state.
In AdS space, this interpretation is somewhat formal because the
boundary is at spatial infinity.  Especially in the case of a
Lorentz signature AdS space,  to have a quantum state flowing in
from {\it spatial} infinity does not correspond to standard
physical intuition.\foot{Nevertheless, this viewpoint has been
adopted in some previous papers, for example \ref\blocks{E.
Witten, ``{\rm AdS/CFT} Correspondence And Topological Field
Theory,'' JHEP {\bf 9812:012} (1998), hep-th/9812012.}.}

\nref\witda{E. Witten, ``Quantum Gravity In de Sitter Space,'' hep-th/01060109.}%
 \nref\stromds{A. Strominger, ``The dS/CFT Correspondence,'' JHEP {\bf 0110:034}
 (2001).}%
  The situation is somewhat different in de Sitter space, with
{\it positive} cosmological constant.  Here the boundary is at
past and future infinity, and it is perfectly natural to
interpret the physical conditions at the boundary as defining an
initial or final state $|i\rangle$ or $\langle f|$, and the path
integral as a matrix element $\langle f|i\rangle$. From this
point of view, any operators that might be inserted in the past
and future induce changes in the initial and final states. Hence
the boundary correlation functions, studied in
\refs{\witda,\stromds,\andetal,\poly}, arise by expanding the
transition matrix element $\langle f|i\rangle$ in ``perturbation
theory'' around the de Sitter invariant state,\foot{Existence of
more than one de Sitter invariant state, as  investigated in
\refs{\andetal,\poly},  presumably means that there are different
possible Hilbert spaces -- inequivalent quantizations of the
field -- as occurs in flat spacetime in the presence of
spontaneous symmetry breaking.}
 a process in which one can extract
part of the information contained in that transition matrix
element. The correlation functions were approached from this
standpoint in \witda.

\bigskip
This work was supported in part by NSF Grant PHY-0070928.  I
would like to thank J. Maldacena, E. Silverstein, and I. R.
Klebanov for discussions. \listrefs
\end